\documentclass[review]{elsarticle}

\usepackage{lineno,hyperref}
\usepackage{graphicx}
\usepackage{float}
\usepackage{amstext}
\usepackage{amsmath}
\usepackage{gensymb}
\usepackage{amssymb}
\usepackage{amsmath}
\usepackage{upgreek}

\journal{Diamond and Related Materials }

\begin{document}

\begin{frontmatter}

\title{Ion Energy Tuning for Enhanced \textit{sp$^{3}$} Carbon Fraction in Electron Cyclotron Resonance Ion Beam Deposited Diamond-Like Carbon Coatings: a Computational and Experimental Approach}

\author{Callum Wiseman $^{a,1}$ \fnref{myfootnote}, Marwa Ben Yaala $^{a,1,2}$ \fnref{myfootnote}, Chalisa Gier $^{a}$, Laurent Marot $^{b}$, Christopher McCormick $^{a}$, Sheila Rowan $^{c}$ and Stuart Reid $^{a}$}

\address{%
$^{a}$ \quad SUPA, Department of Biomedical Engineering, University of Strathclyde, 
Glasgow, United Kingdom;\\
$^{b}$ \quad Department of Physics, University of Basel, Basel, Switzerland; \\
$^{c}$ \quad SUPA, Institute for Gravitational Research, University of Glasgow, Glasgow, United Kingdom;}

\fntext[myfootnote]{These authors contributed equally to this work}
\fntext[myfootnote]{ marwa.ben-yaala@strath.ac.uk}

\begin{abstract}
A novel high-energy electron cyclotron resonance (ECR) ion beam deposition (IBD) technique was used to fabricate DLC films at different ion beam energies. 
The ratios of $sp^2$/$sp^3$ bonding in the DLC coatings were determined by Raman spectroscopy and XPS, with the confirmation of being hydrogen-free due to the lack of photoluminescence (PL) background in the Raman spectra.
The results indicate that the $sp^3$ percentage ranges from 45\% - 85\% for the ECR-IBD fabricated DLC films in this study. Monte-Carlo based SRIM simulation was used to extract the energy and angular distribution of the sputtered particles from the carbon target and correlate it to the highest $sp^3$ fraction in the manufactured ECR-IBD DLCs. This study demonstrates a method of depositing DLC thin films under ambient conditions (room temperature with no post-annealing or additional bias voltage applied) which produces high-$sp^{3}$ coatings (higher than those traditionally reported for other sputtering methods) suitable for applications where high quality DLC coatings are required.

\end{abstract}

\begin{keyword}
\texttt{DLC, ECR-IBD, $sp^{3}$, SRIM, Raman, XPS}
\end{keyword}

\end{frontmatter}


\section{Introduction}
\begin{figure}[H]
\centering
\includegraphics[width=0.5\textwidth]{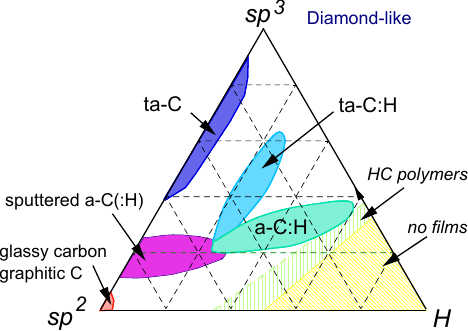}
\caption{Ternary phase diagram of DLC reproduced from \cite{Robertson2011}}
\label{ternary}
\end{figure}

Diamond-like Carbon (DLC) is a classification of amorphous carbon thin films containing some fraction of trigonally and tetrahedrally ($sp^{2}$ and $sp^{3}$, respectively) bonded carbon atoms and may or may not be hydrogenated. In pure $sp^{2}$ carbon (graphite), each C atom has a double bond with three other carbons, whereas in pure $sp^{3}$ carbon (diamond), each C atom has a single bond with four others . Figure \ref{ternary} depicts a phase diagram of $sp^{2}$/$sp^{3}$/H-content and the types of DLC coatings permissible by different combinations of these three properties. 

\noindent DLC coatings are desirable for their varied and interesting properties and their ability to be tuned to meet the unique needs of various applications. They can exhibit many of the properties of crystalline diamond, including high mechanical hardness, chemical inertness, high electrical resistivity, and optical transparency, as well as being a wide band gap semiconductor. Amorphous carbon approaching a 100\% $sp^{2}$ network will have graphitic character while higher fractions of $sp^{3}$ bonding will have more `diamond-like` characteristics. DLC coatings may be deposited using a variety of techniques such as ion beam deposition IBD \cite{Aisenberg1971}; filtered cathodic vacuum arc (FCVA) \cite{chhowalla1997}; RF magnetron sputtering \cite{zambrano2003} and RF plasma enhanced chemical vapour deposition (PECVD)\cite{Varade2014}. As shown in \cite{vetter2014}, each method has its own benefits, drawbacks, and range of achievable film properties.

\noindent Some techniques employed to produce DLC with high $sp^{3}$ content (particularly chemical vapour deposition - CVD) introduce a high amount of hydrogen into the coating due to the use of hydrocarbon precursor gasses. The increased hydrogen content in the coatings may be undesirable for certain applications as it may lead to a decrease in many of the favourable properties that are achieved by the high $sp^{3}$ content, mainly friction and wear \cite{Higuchi2017}.

\noindent This paper presents a room-temperature high-energy sputtering technique for producing hydrogen-free DLC coatings with tuneable $sp^{3}$ content along with a computational study to correlate the sputtering plume energy distribution to the measured coatings composition. The deposition technique here is based on electron cyclotron resonance (ECR)-IBD, a thin-film deposition process previously demonstrated to produce high density hafnium dioxide coatings \cite{Gier2023} and the lowest IR absorption in amorphous silicon thin films \cite{Birney2018}. To the best of the authors’ knowledge, no comprehensive work has previously been undertaken to assess the viability of the ECR-IBD process, or any comparable high-energy sputtering technique, for producing high performance DLC coatings; for example, with a high proportion of $sp^{3}$ bonding relative to other sputtering methods, as presented here.

\section{Experiment}
\subsection{Thin film deposition}
\begin{figure}[H]
    \centering
    \includegraphics[width=.8\textwidth]{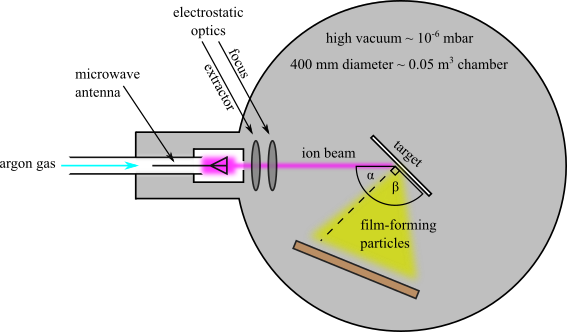}
     \caption{A detailed schematic of the ECR-IBD coating system used in this study.}
    \label{schematic}
\end{figure}

The DLC thin films in this study were produced using a custom-built ECR-IBD system, as shown in Figure \ref{schematic}.

The ECR source (manufactured by Polygon Physics) ionises and accelerates the Ar gas to sputter a target of the desired material, with the sputtered particles contributing to thin film formation. Assuming single ionisation in the argon plasma, we attribute the nominal beam energy to the accelerating voltage applied in the range of 4 - 16 kV for these experiments. Argon is injected into a 1/4$\lambda$ resonant cavity, where it is ionised in the presence of a superimposed AC electric and static magnetic field. At 2.45~GHz, a magnetic field strength of 87.5~mT is required. A beam is then extracted through a series of electrostatic optic-labelled extractors, focus, and ground, as shown in Figure \ref{schematic}. In this study, the potential at the extractor was analogous to the nominal beam energy. The focus was set to 80-90\% of this value, while the ground plate was held at 0~V. The incidence angle, $\alpha$ (normal to the target relative to the ion beam propagation direction), was set to 45$^{\circ}$. The source-target and target-sample distances were kept constant. The substrates coated in this work were either fused SiO$_{2}$ glass or stainless steel 316L. All the depositions were performed at room temperature, and no additional heating was applied to the substrates during the coating process.
Detailed deposition parameters are shown in Table \ref{tab:params}

            \begin{table}[H]
                \centering
                    \begin{tabular}{l r}
                    \textbf{Parameter} & \textbf{Value} \\\hline
                    Target material & HOPG \\
                    Target purity & 99.99\% \\
                    Argon pressure & $3.7 \times 10^{-5}$ to $1.2 \times 10^{-4}$~mbar \\
                    Ion beam voltage & 4-16~kV \\
                    Ion beam current & 100-500~$\mu$A 
                    \end{tabular}
                \caption{Detailed deposition parameters for this study.}
                \label{tab:params}
            \end{table}

\subsection{Characterisation}
SRIM (Stopping and Range of Ions in Matter) simulations were carried out using ion energies across the range of accelerating voltages that the ECR-IBD system was capable of achieving. SRIM is a stochastic code that uses the Monte Carlo approach based on binary collision approximation \cite{ziegler2010}. In this study, SRIM \cite{Ziegler1985} simulation (version 2013) was used to determine the spatial and energy distributions of sputtered atoms from the carbon target. A detailed full-damage cascade TRIM (TRanport of Ions in Matter) calculation was used to record the positions and associated energies of sputtered species from the target. One million Ar ions impinged onto the target material at 45$^{\circ}$ incident angle and with energies of 4~keV, 10~keV and 16~keV. The sputtered target parameters, including the surface binding energy, displacement energy, lattice binding energy, and mass density of the material, were set to the default TRIM values for the carbon target.\\
\newline
Raman spectroscopy/microscopy was used to measure the peak that occurs in carbon materials at around 1580 cm$^{-1}$. Visible Raman spectra were acquired (Renishaw inVia Raman Microscope, Renishaw, Wooton-under-Edge, UK) using an argon laser at an excitation wavelength of 532 nm. Measurements were performed in triplicate, and the spectra were fitted using OriginPro 2021. \\
\newline
X-ray photoelectron spectroscopy (XPS) was used to measure the chemical composition and extract the $sp^{3}$ content. The electron spectrometer was equipped with a hemispherical analyser (Leybold EA10/100 MCD) and a non monochromatised Mg K$\alpha$ X-ray source ($h\nu$ = 1253.6 eV) was used for core level spectroscopy. The binding energy scale was calibrated using the Au 4f7/2 line of a cleaned gold sample at 84.0 eV. The fitting was performed using Unifit 2015 with Voigt peak shapes. The fitting procedure for the core level lines is described in \cite{Eren2011}.

\section{Results and discussion}
\subsection{Raman Spectra of the produced DLC coatings}

\begin{figure}[H]
                \centering
                \includegraphics[width=\textwidth]{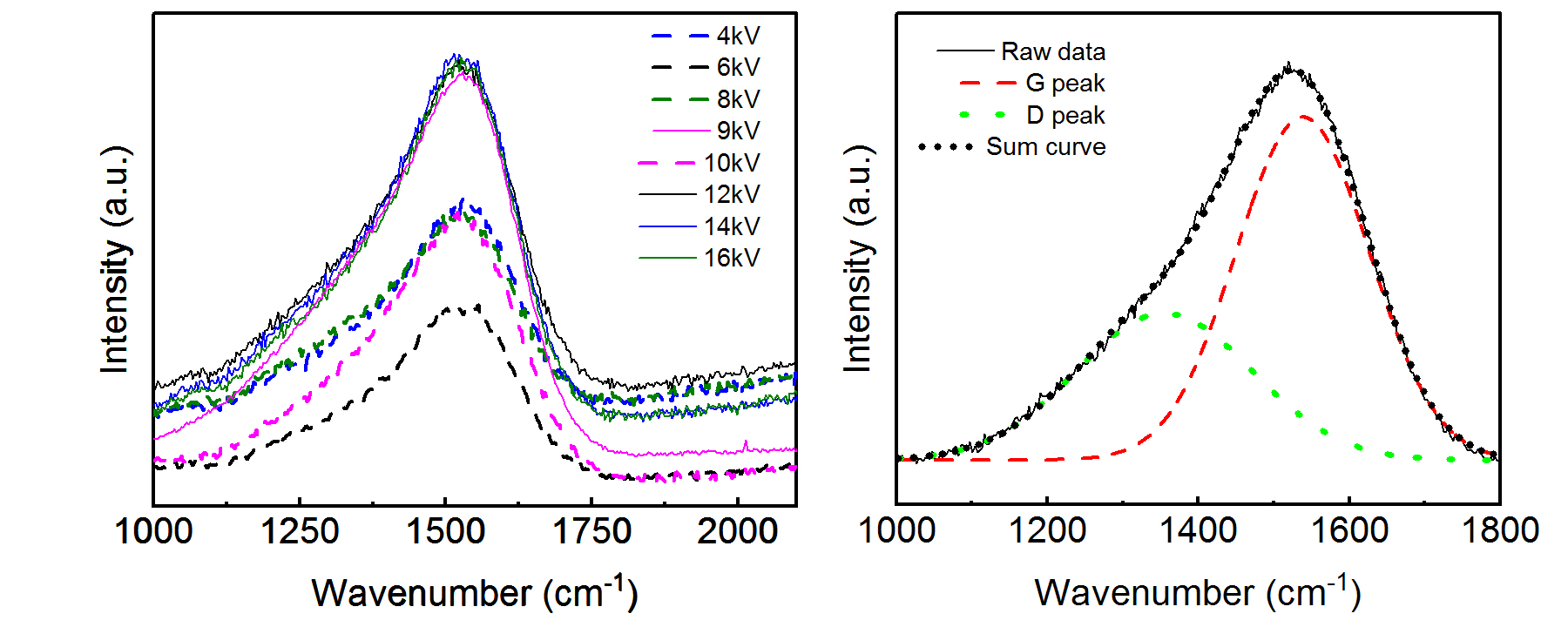}
                \caption{All Raman raw spectra (left) and an example fit (right) to the Raman carbon peak measured on the sample deposited at 10 kV.}
                \label{Raman_fits}
            \end{figure}

Raman spectra for all samples along with their respective fits are shown in figure \ref{Raman_fits}. From the Raman spectra, it can be seen that there is no photoluminescent (PL) background, indicating that the DLCs produced are hydrogen free coatings as PL is mainly due to the hydrogen saturating the non-radiative recombination of the electron–hole pairs in the $sp^2$ bonded clusters \cite{Casiraghi2005}. The C peak in the Raman spectra was deconvoluted into two Gaussian peaks representing the D and G components. Clear differences can be seen in the size and shape of the D and G peaks, as shown in red and green, respectively. The relationships between various parameters of these peaks can be used to deduce structural information about the DLC coatings. The position, width, and intensity of the D and G peaks in the Raman spectra are closely related to the density, size, and structure of the $sp^2$ clusters in DLC \cite{Robertson2002}. Both D and G peaks are due to bond stretching modes in the graphitic structures. The G peak is due to the stretching of carbon bonds either in chains or rings, whereas the D peak is due to the breathing mode of the six-fold aromatic rings. The diamond signature in the DLC referred to in the literature as the T peak (presenting the C-C bond vibration in $sp^3$) can only be seen for UV-excited Raman. For visible excitation, including the current study at 532~nm, only D and G peaks are measurable. In theory, one visible Raman spectrum could correspond to different $sp^3$ contents as the $sp^2$ clustering varies independently of the $sp^3$ bonds so only a combination of a UV Raman and Vis Raman can fully characterise the structure. However, a large number of studies show that the properties of $sp^2$ clusters are in turn closely related to the $sp^3$ content of DLC. In this paper, the the Visible Raman measurements are therefore used as a qualitative estimation of $sp^3$ content of the produced DLC thin films.  Various Raman parameters are often used to infer structural information about DLCs, including the POS(G), I(D)/I(G), and FWHM(G). Specifically, FWHM(G) was used to estimate the $sp^3$ content in the coatings based on the following relationship \cite{Cui2010}:

\begin{equation} \label{eq:cui sp3}
             	sp^3 \; \% = -2.05 + 1.90 \times 10^{-2} = W - 3.01 \times 10^{-5} W^2 \pm 0.08,
            \end{equation}

where $W$ is the FWHM(G) measured at $\lambda$ = 514~nm and $\pm 0.08$ is the standard deviation of the fit of Cui's data to the quadratic model. The following relationship was used to account for the Raman measurements taken at different excitation wavelengths:
            \begin{equation}
            	\mathrm{FWHM(G)_{\textrm{514} \; nm}} = \mathrm{FWHM(G)_{\lambda}} + 0.21 \times (514 - \lambda),
            \end{equation}

where $\lambda$ = 532 nm and FWHM(G) are given in wavenumbers. This equation is based on the dispersion data presented by Ferrari \textit{et al.} \cite{Ferrari2004} with the assumption that FWHM(G) disperses linearly and at approximately the same rate for all types of DLC samples.

\noindent Table \ref{ramantab} presents the $sp^3$\ percentage calculated from the procedure described in equation \ref{eq:cui sp3}. According to this analysis, the majority of samples have $sp^3$ amounts between 40-50\%, with a very slight parabolic dependence going from 48.1\% at 4 kV to 49.9\% at 9 kV and 45.4\% at 16 kV. However, the highest value was 10 kV, with 57.7\% of the carbon fraction being $sp^3$.

            \begin{table}[H]
                \centering
                \begin{tabular}{c c}
                    \textbf{Ion energy (kV)} & \textbf{$sp^3$ estimate from Raman (\%)} \\\hline
                    4   & 48.1  \\
                    6   & 49.3  \\
                    8   & 42.1  \\
                    9   & 49.9  \\
                    10  & 57.7  \\
                    12  & 48.4  \\
                    14  & 47.1  \\
                    16  & 45.4
                \end{tabular}
                \caption{Results of the Raman fitting procedure and $sp^3$ calculation from equation \ref{eq:cui sp3}.}
                \label{ramantab}
            \end{table}

It is worth mentioning here that Cui \cite{Cui2010} model is an empirical model established to fit a specific set of experimental data. This can present a discrepancy for the DLC coatings presented here, and is used for estimating the $sp^3$ content in the coatings. In addition, recent studies have not recommended using Raman spectroscopy to measure the $sp^3$ content of DLCs \cite{Muenz2020, Ba2022}. Therefore, a direct measurement of the composition is performed using XPS to precisely quantify the percentage fractions of carbon bonds in the $sp^2$ and $sp^3$ configurations. Results are presented in the following section.

\subsection{$sp^3$ ratio measured by XPS}
\begin{figure}[H]
                \centering
                \includegraphics[width=.7\textwidth]{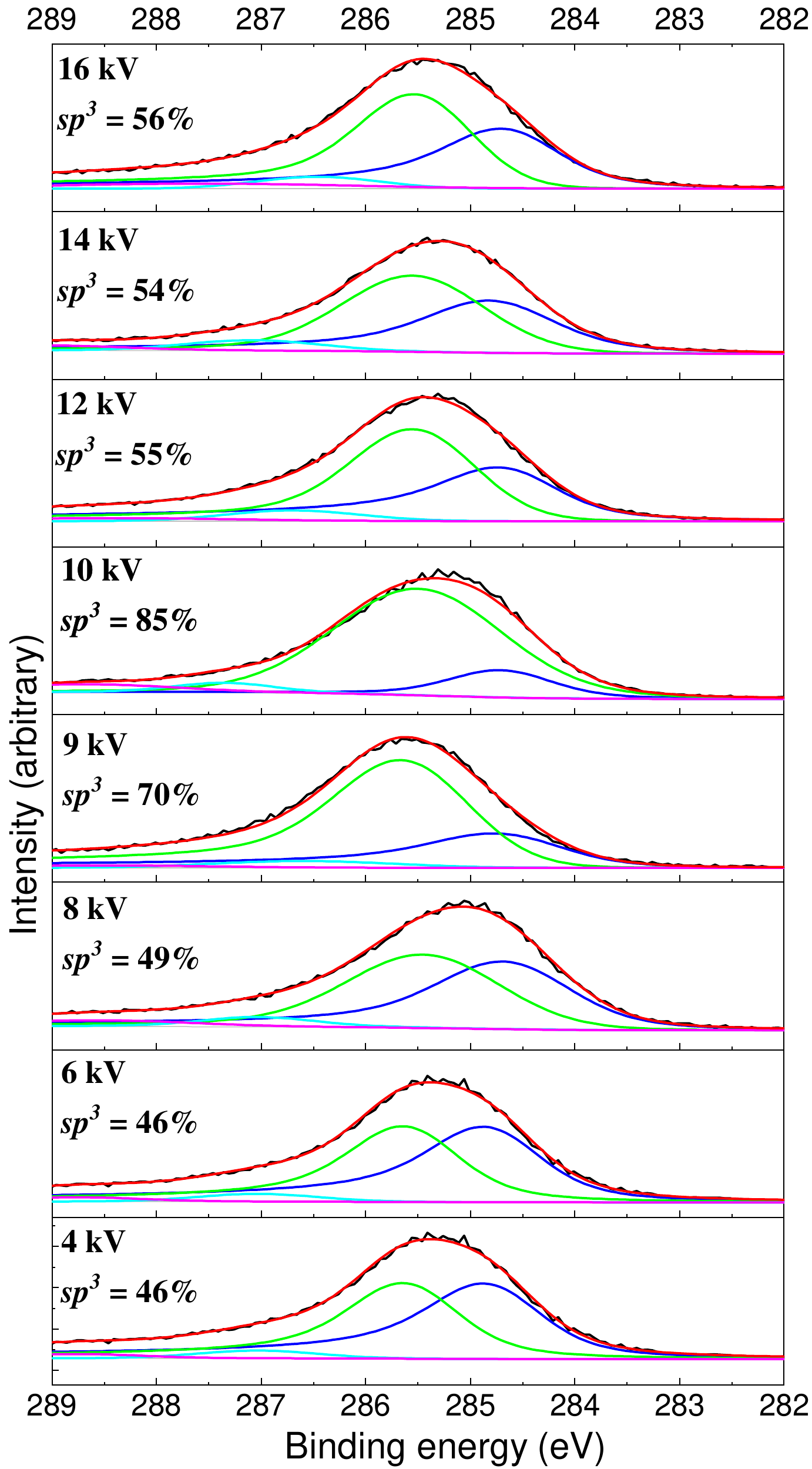}
                \caption{The C1s peak from XPS spectra acquired for samples deposited at different energies. The black curves are the raw measured data; blue, green, cyan, and magenta fitted curves are the individual chemical states; and red curves are the sum curves. The $sp^3$ content is derived from the ratio of the areas of the C-C and C=C peaks fitted in green and blue lines respectively.}
                \label{xps_fits}
        \end{figure}

XPS analysis/fitting of the coatings are shown in figure \ref{xps_fits}. The C1s peak has two components for C=C ($sp^2$) and C-C ($sp^3$) bonded carbons and a 'shoulder' towards higher binding energy for C-O and C=O bonds. Surface oxygen contamination is unavoidable if the samples are exposed to atmosphere after deposition.

\noindent Samples at different ion energies from 4 kV to 16 kV were measured. Shown in figure \ref{xps_fits} are the C1s core level spectra measured for all samples. The C1s core level peak was deconvoluted into four singlets. The peaks at 286.8($\pm0.2$)~eV and 288.4($\pm0.2$)~eV are attributed to C-O and C=O \cite{Lomon2018} while the peaks at 284.7($\pm0.2$)~eV and 285.6 ($\pm0.2$)~eV are assigned to the $sp^2$ and $sp^3$ components of the DLC \cite{Li2018}. Although not shown here, some XPS spectra shifted to higher binding energies during the measurement due to surface charging on the insulating glass substrate. The shift was corrected with respect to the main $sp^2$ and $sp^3$ peaks.
The measurements also revealed the presence of carbon and oxygen. The presence of oxygen on the surface was due to the air exposure of the samples after deposition.

\begin{figure}[H]
                \centering
                \includegraphics{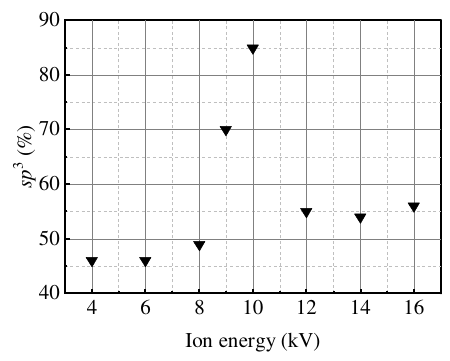}
                \caption{$sp^3$ content estimated from XPS fitting results.}
                \label{xps_scatter}
            \end{figure}

The output of the C1s peak fitting are presented in a scatter plot in figure \ref{xps_scatter}. All samples had $sp^3$ values higher than 45\%, reaching a maximum of 85\% at 10 kV. This value is notably high compared with $sp^{3}$ fraction in DLCs manufactured by other sputtering techniques which generally produce DLC coatings with $sp^{3}$ contents lower than 50\% \cite{vetter2014}. It is important to note that the ECR IBD process requires no bias voltage on the substrate,which provides more flexibility in the choice of substrate (insulating, temperature sensitive, etc.) to produce high $sp^{3}$ DLCs for different applications.

\noindent The large disparity between the $sp^3$ fractions estimated from the Raman and measured by XPS is found where raman gives lower estimates overall than XPS. Raman is only a qualitative method, it does not measure the exact $sp^3$ content however the position of the G peak indicates a ta-C phase (see Robertson \cite{Robertson2002} for POS(G) at 513 nm). Both XPS and Raman show that $sp^3$ depends on the energy of the ions sputtering the target and, consequently, on the energy of the sputtered atoms. 

The growth mechanism of DLC films is best understood by consideration of subplantation of incident particles \cite{lifshitz1989}. Up to a certain value, incoming film-forming particles will not have enough energy to cause the densification required to produce a high proportion of $sp^{3}$ bonds, whereas at higher energies, too much thermal dissipation will occur, resulting in a lower fraction of $sp^{3}$ bonds being formed. Robertson \cite{Robertson2002} suggests that carbon ions with energies of around 100 eV will produce high $sp^{3}$ content DLCs, although Zia \cite{zia2022} notes that value may vary with deposition technique: some producing high $sp^{3}$ DLCs at 140 eV \cite{fallon1993}; others producing adequate $sp^{3}$ content at as low as 30 eV \cite{mckenzie1991}. Therefore, it is important to estimate the energies of the sputtered particles that contribute to film formation. When using a process such as FCVA, a common technique for producing DLC with a high proportion of $sp^3$ bonding, the energy of the film-forming particles is the same as the bias voltage applied to the substrate, so is directly known. However, in the ECR IBD process, the energy of the Ar$^+$ ions are known, unlike the energy of the film-forming particles sputtered from the target. Hence, SRIM simulation was used to calculate the energy distribution of the film-forming particles.

 \subsection{Sputtered particles energy: simulation results}
\begin{figure}[H]
                \centering
                \includegraphics[width=\textwidth]{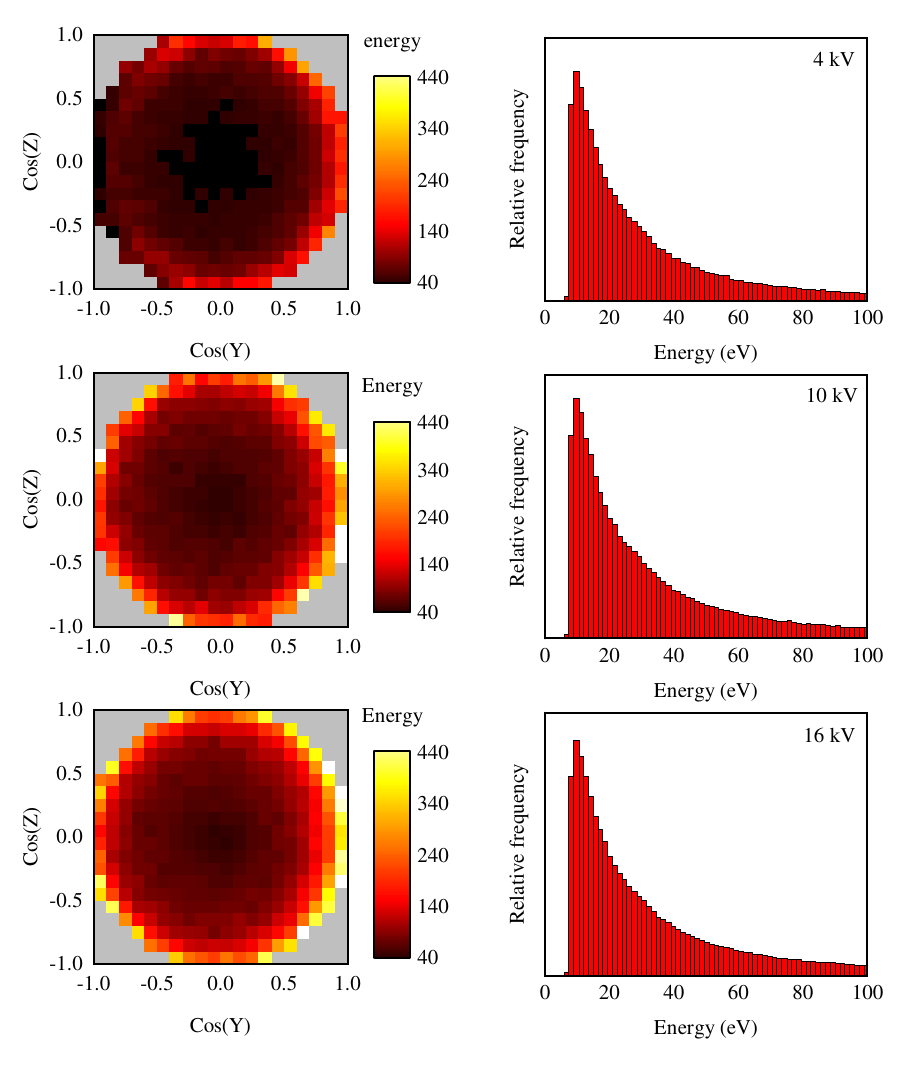}
                \caption{Heatmaps (left) from SRIM simulation of the energy/angular distribution of film forming C atoms sputtered using 4, 10, and 16 kV (top-bottom) ion beam and the flux (right) or energy distribution of sputtered atoms against their relative frequency.}
                \label{SRIMHM}
            \end{figure}

The angular energy distribution of the sputtered particles from the carbon target was simulated using the Monte Carlo software SRIM \cite{ziegler2010} for different ion beam energies.
Figure \ref{SRIMHM} shows the energy and angular distribution of the film forming particle and their relative frequency. High flux of low energy particles are found in the middle of the sputtering plume, while a lower flux but higher energy particles are sputtered towards higher emission angles.

            \begin{figure}[H]
                \centering
                \includegraphics[width=\textwidth]{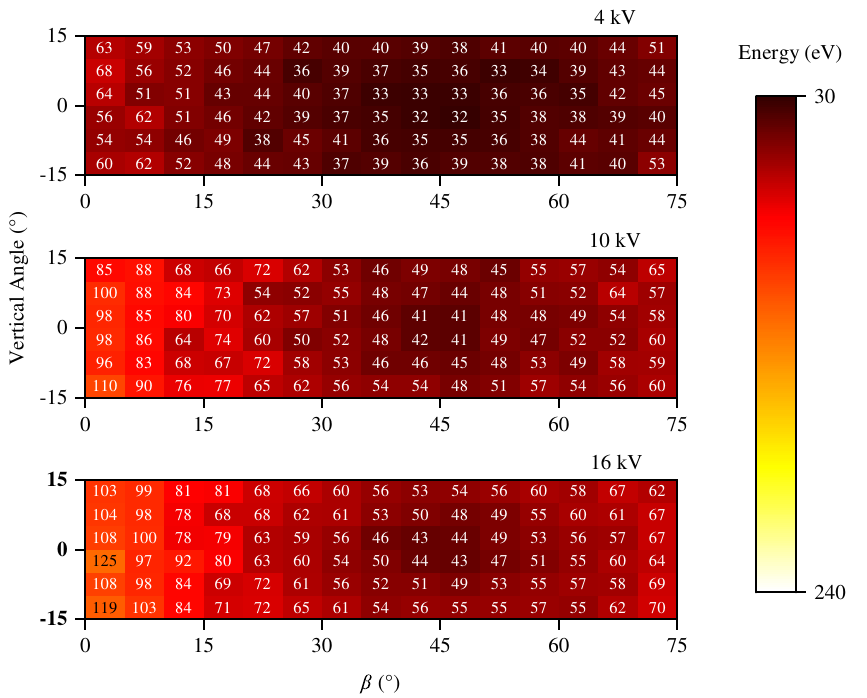}
                \caption{Sections of the SRIM simulations with 4 kV (top), 10 kV (middle), and 16 kV (bottom) ion beams. The energies seen by the substrates positioned at various positions within the plume are annotated in white in the figure.}
                \label{srimpos}
            \end{figure}

Figure \ref{srimpos} presents a plane section of the sputtering plume energy distribution representing the energies seen by the substrates positioned at various positions within the plume. The C atoms were sputtered from the target with energies of approximately 30-60~eV at 4~kV, 40-100~eV at 10~kV, and even as high as 125~eV at 16~kV. High $sp^3$ content is correlated with high density in DLC films, with pure diamond having a density of 3.52 g/cm$^3$. As the density of the deposited films approaches this value, the $sp^3$ fraction also increases. High density is achieved when incoming atoms have an optimum energy, which is high enough to be sub-planted in the deposited film but low enough not to cause thermal dissipation in the film (Robertson 2011). If the incoming atom energy is too low, it will stop at the surface, resulting in low density; if the energy is too high, some of that energy will be dissipated thermally in the film, resulting in the dispersion of already implanted atoms, in addition to thermally driven structural changes, for example, localised phase transitions and potential modifications to internal stresses. According to the XPS results presented in the previous section, the optimum ion energy for the highest $sp^3$ fraction is at 10~kV which corresponds to 41~eV sputtered atom energy in the middle of the plume, where the flux is the highest.

\section{Conclusion}
Hydrogen-free DLC coatings were deposited at room temperature using the ECR-IBD process. Raman spectra of the coatings were analysed and fitted and the fitting parameters were used to estimate the $sp^3$ ratio in the deposited films. Direct quantitative measurements were performed using XPS. The impact of the energy of both the ion beam and the sputtered particle on the structure of the manufactured coatings was investigated. The results show that the ECR-IBD technique produces DLC thin films with $sp^{3}$ content up to 85\%, surpassing the values previously reported for other sputtering techniques. 

\section{Acknowledgments}
We are grateful for financial support from STFC (ST/V005642/1, ST/W005778/1, and ST/S001832/1) and the University of Strathclyde. S.R. is supported by a Royal Society Industry Fellowship (INF$\backslash$R1$\backslash$201072). We are grateful to the National Manufacturing Institute for Scotland (NMIS) for support, and we thank our colleagues within SUPA for their interest in this work.

\bibliography{energysp3}

\end{document}